\documentclass[fleqn,11pt]{wlscirep}
\usepackage[utf8]{inputenc}
\usepackage[T1]{fontenc}
\usepackage{bm}
\usepackage{amsmath}
\usepackage[normalem]{ulem} % Put this in the preamble
\usepackage{multicol}
\usepackage{multirow}
\usepackage[most]{tcolorbox} % For shaded boxes
\tcbuselibrary{breakable}
\usepackage{graphicx} % For including images
\usepackage{lipsum} % For dummy text
\usepackage{placeins}
\usepackage{afterpage}
\usepackage[justification=justified]{caption}
\tcbset{
    graybox/.style={
        colback=gray!10, % Light gray background
        colframe=gray!50, % Gray border
        boxrule=0.5mm, % Border thickness
        arc=2mm, % Rounded corners
        left=2mm, % Left padding
        right=1mm, % Right padding
        top=1mm, % Top padding
        bottom=1mm, % Bottom padding
        width=1\linewidth, % Ensure the box spans the text width
        boxsep=1mm, % Padding around content
    }
}
\title{Photonic Ising machines toward and beyond a million spins}
\author[1,2$\dag$]{A. Aadhi}
\author[1$\dag$]{Nayem Al-Kayed}
\author[1]{Tristan Austin}
\author[3]{Charles St-Arnault}
\author[1]{Nir Rotenberg}
\author[4]{Hitesh Ballani}
\author[3]{David V. Plant}
\author[5,6]{Edwin Ng}
\author[5,6]{Yoshihisa Yamamoto}
\author[7*]{Peter L. McMahon}
\author[1,8*]{Bhavin J. Shastri}

\affil[1]{Centre for Nanophotonics, Department of Physics, Engineering Physics and Astronomy, Queen’s University, Kingston, ON K7L 3N6, Canada}
\affil[2]{Optics and Photonics Centre, Indian Institute of Technology Delhi, New Delhi, Delhi 110016, India}
\affil[3]{Department of Electrical and Computer Engineering, McGill University, Montreal, QC H3A 2A7, Canada} 
\affil[4]{Microsoft Research, Cambridge, CB1 2FB, UK}
\affil[5]{Physics \& Informatics Laboratories, NTT Research, Inc., Sunnyvale, California 94085, USA}
\affil[6]{E. L. Ginzton Laboratory, Stanford University, Stanford 94305, CA, USA}
\affil[7]{School of Applied and Engineering Physics, Cornell University, Ithaca, New York 14853, USA}
\affil[8]{Smith Engineering, Department of Electrical and Computer Engineering, Queen’s University, Kingston, ON K7L 3N6, Canada}
\affil[*]{Corresponding authors Email: pmcmahon@cornell.edu, shastri@ieee.org}
\affil[$\dag$]{These authors contributed equally to this work.}

\date{}
\begin{abstract}
Combinatorial optimization problems are central to many challenges in logistics, finance, engineering, and the life sciences, yet they remain among the most computationally demanding. Many of these problems can be mapped onto the Ising model, in which binary spins interact through a network of couplings, and solutions correspond to low-energy, ideally ground-state, spin configurations. Photonic Ising machines have the potential to be fast and energy-efficient heuristic solvers of optimization problems by leveraging the low latency, high bandwidth, and inherent parallelism of optics. However, current photonic implementations remain limited in scalability, connectivity, reconfigurability, and time-to-solution, preventing their use in many practical applications. In this perspective, we examine the current landscape of photonic Ising machines, discuss the challenges and limitations of existing platforms, and identify the scientific and technological advances needed to realize large-scale systems. These developments could establish photonic Ising machines as useful hardware platforms for practical optimization.
\end{abstract}
\begin{document}
\maketitle
\vspace{-18pt}
\noindent \textbf{\large Key points}%  
\begin{itemize}
\item Combinatorial optimization problems arise across drug discovery, materials engineering, wireless communications, logistics, and finance. However, state-of-the-art computing architectures often struggle to find optimal solutions within practical time and energy budgets for many real-world instances. 
\item Physics-based computing platforms have emerged as promising approaches for combinatorial optimization. However, achieving practical large-scale implementations will require advances in architecture and hardware-algorithm co-design. 
\item Scalability, connectivity, time-to-solution, and solution accuracy remain major challenges across Ising machine platforms. Photonic technologies offer distinct advantages, including low latency, high bandwidth, inherent parallelism, coherence, and room-temperature operation. 
\item All-optical and optoelectronic photonic platforms using spatial, temporal, wavelength, or hybrid multiplexing are promising candidates for next-generation scalable photonic Ising machines. 
\item Many real-world combinatorial optimization problems require millions of variables when mapped to the Ising formalism. To address such problems natively, photonic Ising machines will need to support millions of spins. Several approaches could plausibly reach this scale and beyond.

\end{itemize}

%%%%%% Main Text %%%%%%
\section{Introduction}
\label{sec:intro}
Solving large-scale combinatorial optimization problems, even approximately, is computationally intensive because the solution space grows exponentially with problem size \cite{imitriou1998,hochba1997, cook2011traveling}. Such problems--many of which are nondeterministic polynomial-time hard (NP-hard)-- arise in many fields, including finance \cite{mantegna1999introduction}, chemistry \cite{mizuno2024finding}, biology \cite{ liu2000dna}, drug design \cite{kitchen2004docking}, circuit design \cite{barahona1988application}, artificial intelligence (AI) \cite{smith1999neural} and robotics \cite{chakraa2023optimization}. In principle, many of these complex problems can be mapped to the century-old Ising model\cite{ising1924beitrag}, a network of tiny magnets called “spins,” and addressed with an Ising machine \cite{strachan2021fast}. Ising machines have attracted significant attention for their ability to heuristically find near-optimal, and in some cases, optimal solutions to problems faster than conventional algorithms on conventional hardware \cite{mohseni2022ising}.  The Ising model, originally developed in statistical mechanics, provides a mathematical framework for describing magnetization, phase transitions, and critical behavior through energy minimization \cite{ising1924beitrag}.  The Ising Hamiltonian defines the energy of the Ising model $E = -\sum_{\langle i,j \rangle}^{N} J_{ij} s_i s_j - \sum_{i}^{N} h_i s_i  $ where \textbf{\(J_{ij}\)} is the coupling strength between spins $i$ and $j$, \textbf{\(s_i \in \{+1,-1\}\)} represents the binary spin states at node \(i\) and \textbf{\(h_i\)} is an external magnetic field. The first term captures spin-spin interactions, the second term describes the interaction of each spin with the external field. By encoding an optimization problem into the coupling matrix and fields, the solution corresponds to a low-energy--ideally ground state configuration of the Ising Hamiltonian. A wide range of physical platforms have been proposed and demonstrated as Ising machines, including superconducting circuits\cite{johnson2011quantum,razmkhah2024josephson}, trapped ions \cite{kim2010quantum}, electromechanical oscillators \cite{mahboob2016electromechanical}, complementary metal-oxide-semiconductor (CMOS)-based electronic architectures \cite{yamaoka201520k,tatsumura2021scaling}, memristors\cite{cai2020power, jiang2023efficient, hizzani2023memristor}, polaritons\cite{kalinin2018simulating, kalinin2020}, spintronics systems\cite{ albertsson2021, houshang2022, sutton2017}, and photonics ~\cite{gao2024photonic}. Photonic Ising machines  are particularly attractive because they leverage high bandwidth, low latency, low loss, and room temperature operation \cite{gao2024photonic}. Several photonic architectures based on spatial, temporal, or wavelength multiplexing have been demonstrated across various hardware platforms, with the goal of improving  scalability, reconfigurability, and computational speed (see Table~\ref{tab:1}). Despite this progress, photonic Ising machines face three intertwined challenges: scalability (the ability to realize large spin systems), connectivity and precision (the ability to program accurate couplings); and time complexity (the ability to achieve near-optimal solutions within a reasonable time). 

\textcolor{black}{Present implementations remain far below the scale required for practical applications such as biomolecular simulations, drug discovery, protein folding, channel allocation, data-centre scheduling, traffic flow management and brain network modelling \cite{mizuno2024finding, al2025programmable, yao2025enhanced,bao2023ising, lee2026fundamental}. Therefore, the requirement of large-scale photonic Ising machines is substantial.  For example, the travelling salesman problem, a canonical problem with applications in routing, airline scheduling, and logistics, requires $N^{2}$ spins to map \textit{N} cities, so optimizing routes among 1,000 cities requires approximately one million \cite{si2024energy}. Protein folding with applications to drug discovery presents a similar challenge: a binary encoding of the two-dimensional hydrophobic–polar (HP) lattice model for a protein with \textit{N} amino acids on a $L^{2}$ lattice, the number of spin variables scales as $L\times N^{2}/2$, so identifying the structures of a 200-amino-acid protein on a $100\times100$ lattice requires millions of spins \cite{al2025programmable, irback2022folding}.  These examples highlight the need for hardware that supports millions of programmable parameters while simultaneously addressing scalability, reconfigurability, and time complexity.} Furthermore, real-world problems often require non-trivial mapping formulations that differ from those used in benchmarking \cite{kalinin2025analog}. \textcolor{black}{Practical Ising formulations support linear or nonlinear constraints, binary or continuous variables, high bit precision and mapping overhead with auxiliary variables but increase the required number of spins. } The current CMOS architectures offer monolithic integration, high-bit precision and clock rates ranging from hundreds of MHz to GHz, but their scalability and connectivity remain limited. On the other hand, multi-core CMOS processors overcome these limitations and support large-scale computation through parallel processing.\textcolor{black}{For example, a 25× NVIDIA H100 GPU cluster can potentially implement a million variable MVM operations. However, it consumes 10s of kW of GPU power and costs several million USD \cite{nvidiaNVIDIAH100Tensor}. In addition, distributed synchronization latency, multi-terabyte storage, communication overhead, and cooling and operational power requirements are the primary challenges in deploying multi-GPU clusters. Further extending the architecture to implement a million-node Ising machine requires additional computational cost for spin updates and signal processing, in addition to long convergence times and high power consumption associated with simulated annealing. On the other hand, the recently demonstrated multi-board FPGA-based CMOS annealing processor implements a massive scale of 1.3 million Ising spins to solve combinatorial optimization problems \cite{yamamoto20211}. } 

\begingroup
\fontsize{10}{12}\selectfont
\begin{tcolorbox}[graybox]
\textbf{Box 1 | Photonic Ising machines: concept and principle} 

Photonic Ising machines typically comprise three core elements: spin encoding, implementation of spin-spin interactions through a coupling matrix, and an optimization algorithm. Their physical implementations are commonly classified by architecture, encoding method, and algorithmic framework.  
\begin{center}
\includegraphics[]{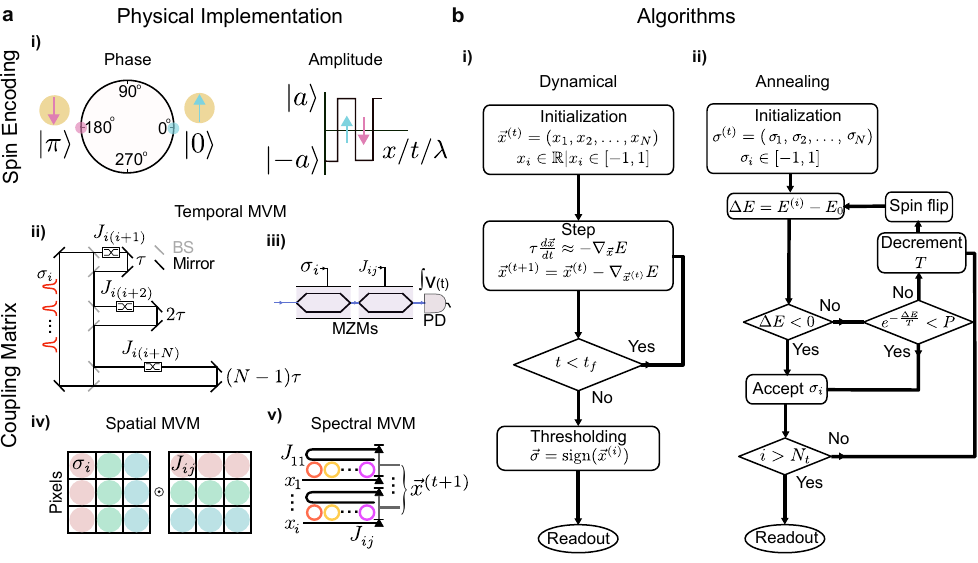}
\label{fig:2}

\end{center}
 \begin{multicols}{2}
\textit{Encoding}:
In photonic implementations, Ising spins are commonly encoded in the phase, amplitude or intensity of the continuous-wave or pulsed optical signals. Figure a (i) illustrates spin encoding using phase and amplitude. Phase or amplitude encoding can be implemented in different degrees of freedom of a light, including space (\textit{x}), time (\textit{t}) or wavelength ($\lambda$). Spin states can be represented using either discrete or analog variables, and the choice of encoding scheme directly influences system performance \cite{zhang2024review}. Analog encoding represents spin states as continuous values rather than restricting them to binary states. These states can be encoded through modulation of optical amplitude, phase, or intensity; through optical nonlinear processes; or by exploiting optoelectronic nonlinearities.
\\
\textit{Coupling matrix}:
Spin-spin interactions in the Ising model can be implemented through matrix-vector multiplications (MVMs). Iterative MVM operations update the spin state towards a low-energy configuration. Photonic MVMs have been explored using several multiplexing strategies, including spatial multiplexing based on plane-wave diffraction and Mach–Zehnder interferometer (MZI) meshes; wavelength multiplexing using wavelength-selective filters such as tunable microring resonator weight banks and crossbar arrays; and time multiplexing using delay lines and cascaded electro-optic modulators \cite{zhou2022photonic}. Figure a (ii)-(v) shows representative implementations of photonic MVMs. For all-to-all connected spin configurations, the interaction matrix contains $N^2$ non-zero elements, creating hardware challenges associated with routing, memory and control. Sparsification techniques can restructure the connectivity and reduce the number of non-zero elements, enabling more scalable computation within hardware and memory constraints. Photonic MVMs offer key advantages, including high bandwidth, low latency, and reconfigurability. 
\\
\textit{Algorithms}: 
Classical Ising algorithms can be broadly grouped into two classes: dynamical system approaches and simulated annealing. Dynamical systems use the natural evolution of a physical system to search the Ising energy landscape. In such systems, physical annealing can guide the dynamics towards low-energy, and ideally ground-state, configurations of the Hamiltonian. Coupled injection-locked lasers and OPO-based coherent Ising machines are examples of dynamical systems that implement this form of physical annealing. Simulated annealing, by contrast, begins with a randomly initialized spin configuration and iteratively updates the spins by evaluating changes in the Hamiltonian, gradually converging towards low-energy configurations. Simulated annealing can use several spin-update strategies, including Monte Carlo methods, parallel tempering, population annealing, and hybrid simulated annealing \cite{aramon2019physics}. Figure b (i)-(ii) shows flowcharts of these two main algorithmic classes. The choice of algorithm affects convergence rate and computational time \cite{lucas2014ising}. 
\end{multicols}

\end{tcolorbox}
\endgroup

\begin{figure}[htbp]
    \centering
     \includegraphics[height=6.3in]{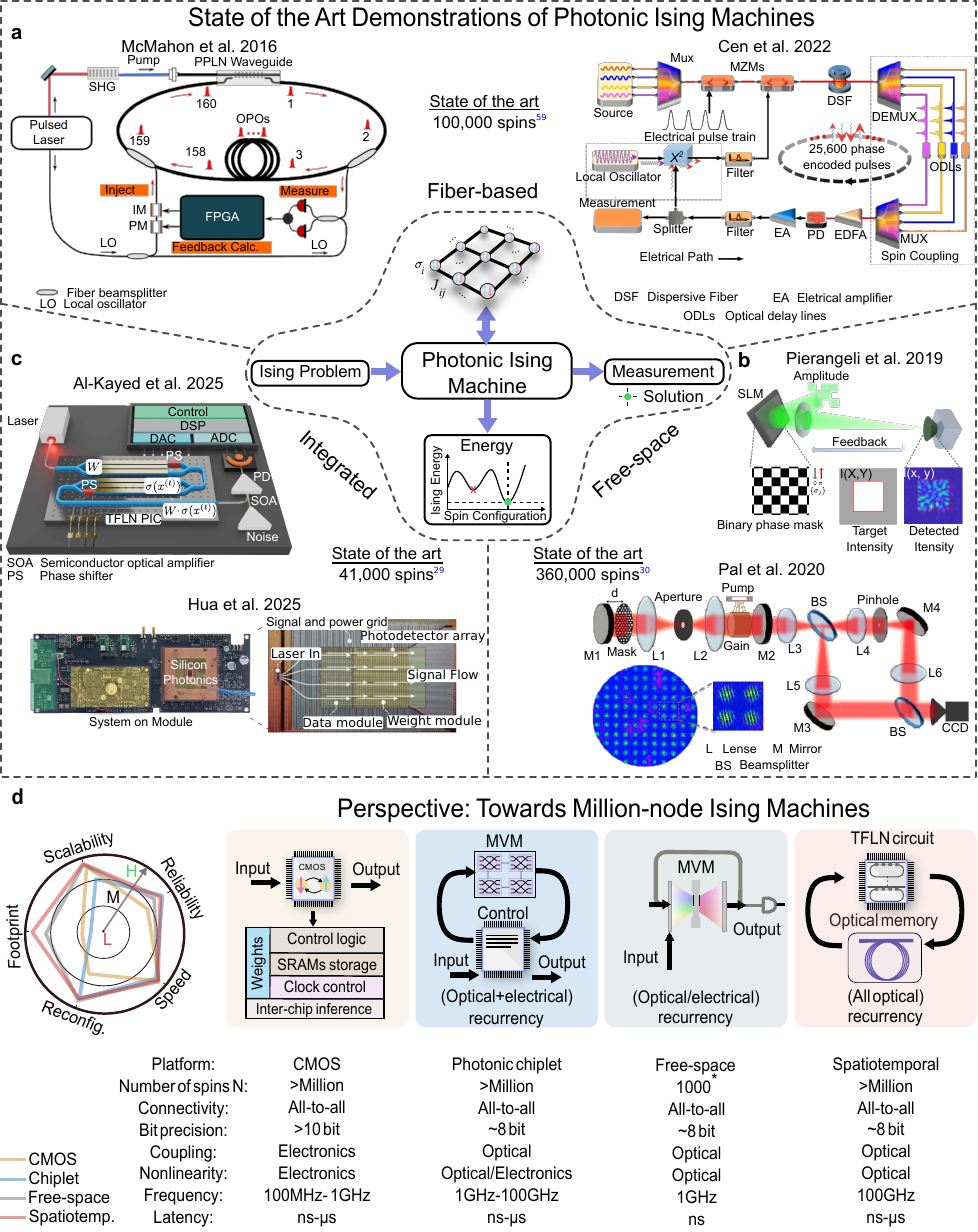}
    \caption{\textbf{Recent progress and the roadmaps toward million-node Ising machines:} Photonic Ising machines have been implemented using various approaches. The figure represents the fiber-based, free-space, and integrated photonic architectures. \textbf{a)} Shows the implementation of two different fiber-based CIMs using time-division-multiplexed OPOs with measurement and feedback signals \cite{mcmahon2016fully} and optoelectronic parametric oscillators \cite{cen2022large}. \textbf{b)} Represents the large-scale spatially multiplexed free-space Ising machine employing spatial light modulation to encode spins \cite{pierangeli2019large} and the implementation of XY-photonic Ising machines using a dissipative coupled laser system \cite{ pal2020rapid}. \textbf{c)} A hybrid integrated photonic Ising accelerator with a co-integrated electronic chip to enable ultralow computation latency \cite{hua2025}, and the on-chip optoelectronic photonic Ising machine employing TFLN modulators \cite{al2025programmable}. \textcolor{black}{\textbf{d)} Roadmap to million-node Ising machines}. The spider chat provides a qualitative illustration of the key advantages of different architectures. The arrow points from low (L) to medium (M) and high (H) values. Typically, CMOS architectures operate at clock rates of 100 MHz to a few GHz, offer large scalability and low latency, but have limited connectivity~\cite{takemoto20214,sharma2022increasing}. Photonic implementations, including chiplet, free-space, and all-optical coherent spatiotemporal architectures, enable scalable, high-speed MVM and offer high reconfigurability~\cite{al2025programmable,calvanese2021all,wang2022optical,takesue2019large}. The $^{*}$  symbol represents an all-to-all connected interaction implemented using the fan-in/fan-out configuration, however, by exploiting Fourier-domain multiplication or multiple free-space modules, it can be scaled to support a million spins. Panel \textbf{a} adopted with permission from ref. \cite{mcmahon2016fully}, AAAS and ref. \cite{cen2022large}, Springer Nature. Panel \textbf{b} adopted with permission from ref. \cite{ pierangeli2019large,pal2020rapid}, APS. Panel \textbf{c} adopted with permission from ref. \cite{ hua2025,al2025programmable }, Springer Nature.}
   
    \label{fig:1}
\end{figure}

\textcolor{black}{The architecture achieves large scalability by integrating 9 FPGA boards to handle 16000 spins/chip. The architecture offers computational advantages, including faster annealing and higher accuracy, compared with CPU-based technologies. However, at large scales, significant challenges arise from the need for large storage capacity, complex routing for interactions, and control circuits for readout and periodic spin updates. In addition, communication overhead dominates computation in multi-board systems. Recently, various analog, digital and mixed-signal Ising machines have been demonstrated using CMOS coupled oscillator systems \cite{ yue2024scalable, cilasun2025coupled,maher2024cmos}, however, the requirement of a large number of spins, all-to-all connectivity, and their interconnections at the chip scale encounters fundamental design and fabrication limits.} On the other hand, photonic architectures implement reconfigurable spin-spin coupling and optical and/or feedback, are easy to design, support large-scale systems at high operational bandwidth with no resistive heating, which can potentially obtain approximate solutions, low latency, high solution accuracy and improved energy efficiency.  Inspired by modern CMOS multicore processors, multicore photonic architectures offer a promising route to fast and reconfigurable Ising processors at a million-scale. Emerging approaches, such as integrated photonic chiplets, all-optical, analog optoelectronic, and all-optical coherent spatiotemporal processors could provide viable paths to million-node systems. Here, we survey existing photonic Ising machine architectures, discuss the major challenges to scalability, and outline feasible technologies for scaling beyond one million spins.

  \section{Recent progress on photonic Ising machines}
The past decade has seen growing interest in photonic Ising machines as scalable, low-latency, stable, and energy-efficient hardware platforms. This interest has driven the exploration of a wide range of photonic Ising architectures. Box 1 summarizes the physical implementation of spin encoding, MVMs, and the main algorithmic frameworks used in photonic Ising machines. Existing approaches span multiple hardware platforms, including free-space, fiber-based, and integrated photonic systems); leverage diverse multiplexing schemes, including time, space, wavelength, and hybrid multiplexing; and employ several algorithmic strategies, including simulated annealing, momentum-based schedules, simulated bifurcation, Hopfield-Tank algorithm, parallel tempering, and related methods \cite{zhang2022,zhang2024review}. 

A generic photonic Ising machine maps problem variables to spins, implements spin–spin interactions via MVMs, and applies iterative feedback to drive the system towards low-energy configurations. Given the diversity of photonic implementations, these systems can be classified in several ways. Here, we categorize photonic Ising machines into three main classes based on hardware implementation: fiber-based, free-space, and integrated photonic architectures. Although all three classes share the objective of minimizing the Ising Hamiltonian to obtain optimal spin configuration (the desired solution), each offers distinct trade-offs in scalability, processing speed, energy efficiency, stability, solution accuracy, and physical footprint. Table 1 summarizes recent advances in state-of-the-art photonic Ising machines.

\subsection{Fiber-based photonic Ising machines}
In fiber-based architectures, fiber-optic components are used for spin generation, spin storage, and/or MVMs \cite{ inagaki2016large, mcmahon2016fully, cen2022large}. In most approaches, spin encoding and updates are realized either through nonlinear wave-mixing processes in degenerate optical parametric oscillators (DOPOs) or through electro-optic modulators in electro-optic oscillator configurations \cite{takesue201610, bohm2021order}. Long-fiber feedback loops enable storage of many time-multiplexed spins (>10k), thereby partially mitigating scalability limitations. In DOPO-based systems, the down-converted light exhibits discrete phase values of either 0 or $\pi$. By contrast, electro-optic oscillators, assisted by MZI modulator nonlinearities, yield discrete amplitude states, typically +\textit{a} or -\textit{a}, which emulate binary spins when operated above the bifurcation point.

Early DOPO-based architectures implemented spin–spin interactions by incorporating optical delay lines into the feedback loop \cite{inagaki2016large}. Although the number of spins is determined by the fiber length and pump-pulse repetition rate, the requirement of \textit{N}-1 delay lines to implement spin–spin interactions severely limits scalability. To address this limitation, measurement–feedback architectures replace optical delay lines with field-programmable gate arrays (FPGAs). In these systems, the interaction matrix is implemented electronically on the FPGA, and the resulting feedback signal is optically reinjected via electro-optic modulation (see Fig.~\ref{fig:1}a) \cite{mcmahon2016fully, honjo2021100, wei2026versatile}. However, as the fiber length increases to accommodate more spins, particularly beyond 100,000, system instabilities become more severe, effectively limiting practical fiber lengths to a few kilometers, typically less than 5 km \cite{honjo2021100}. 

More recent fiber-based optoelectronic oscillators offer improved stability, electronic co-integration, and scalability. In these systems, the electro-optic modulators encode the binary spin values, whereas spin–spin interactions are implemented using FPGAs \cite{bohm2019poor, bohm2021order}. A primary challenge is the latency associated with DAC/ADC data upload and download, and repeated optical-electronic-optical conversion. In both DOPO-based and optoelectronic fibre platforms (see Fig.~\ref{fig:1}a), optimization problems are mapped onto the Ising Hamiltonian, and the iteratively evolving output optical phases or amplitudes are driven towards low-energy spin configurations. Additional details on the fiber-based architectures, including state-of-the-art coherent Ising machines \cite{honjo2021100}, Kerr-nonlinearity-based Ising machines \cite{jin2025kerr, quinn2026coherent} and optoelectronic parametric oscillators \cite{hu2026programmable}, are provided in Supplementary Section 1.1.

\subsection{Free-space photonic Ising machines} 
Free-space photonic Ising architectures have attracted significant attention because they offers spatial multiplexing, massive parallelism, scalability, and low loss. In a typical free-space architecture, spin states are encoded as binary phase values (0 and $\pi$) on an optical wavefront or as a spatial intensity profile using a spatial light modulator (SLM) \cite{pierangeli2020adiabatic, kalinin2025analog}.This flexible spatial-multiplexing configuration enables free-space systems to process tens of thousands of spins in parallel (see Fig.~\ref{fig:1}b). Spin-spin interactions can be implemented either through free-space propagation itself or by incorporating an explicit MVM stage in the optical path. 

Using the large pixel counts available in SLMs, free-space photonic Ising machines have been demonstrated at scales exceeding \textit{N} > 10,000 spins spins with all-to-all connectivity. However, the relatively slow response time of SLMs remains a key limitation, leading to long iteration times ranging from milliseconds to seconds. To improve reconfigurability, solution quality, and time-to-solution, several approaches have recently been explored, including multimode fiber-based systems, higher-order interactions enabled by optical nonlinearities, and wavelength-multiplexing techniques \cite{babaeian2019single, ouyang2022demand, pal2020rapid, ye2023photonic, kumar2020large, WDM_Li2023}. 

More recently, analog electronics feedback combined with fast input-spin encoding using intensity-controlled laser arrays has emerged as a promising route to mitigate the SLM switching bottleneck\cite{kalinin2025analog}. Furthermore, the advantage of free-space architecture is that they naturally map onto two-dimensional spin-glass systems, in which spins interact randomly, enabling efficient solutions to problems involving frustration and a complex energy landscape. As a result, several spin-glass models, including higher-order Ising models and XY spin glasses, have been explored for complex benchmark problems\cite{pierangeli2020adiabatic,pierangeli2021scalable, ouyang2024programmable,fang2021experimental}. 

Gain-based free-space systems offer another route to combinatorial optimization, in which the search dynamics are governed by gain and loss in driven–dissipative systems. Spatially multiplexed laser arrays have been shown to efficiently reach low-energy, and in some cases ground-state, configurations of the Ising Hamiltonian \cite{nixon2013observing, pal2020rapid, gershenzon2020exact}. Further details on free-space photonic Ising architectures are provided in Supplementary Section 1.2.

\subsection{Integrated photonic Ising machines} 
In recent years, substantial effort has been devoted to integrated photonic Ising machines because of their compact footprint, CMOS compatibility, improved stability, and low latency. The central challenge for integrated photonic architecture is scalability: large problem sizes require integrating many optical components onto a single chip. MZI mesh arrays, MZI modulators, and microring resonator arrays are commonly used as MVM cores in integrated photonic Ising machines (see Figure~\ref{fig:1}c)\cite{wu2025monolithically,liu2025chip, al2025programmable,rausell2025ising}. Spin states are typically encoded via amplitude or phase modulation, whereas reconfigurable spin–spin interactions are implemented using optical or optoelectronic MVM cores based on silicon photonics or thin-film lithium niobate (TFLN). Analog or digital electronic feedback circuits then provide iterative spin updates \cite{prabhu2020accelerating, li2024scalable}. 

Among these approaches, MZI meshes are widely used integrated linear-optical systems that can implement universal unitary transformations. The Reck and Clements meshes are the most common architectures for realizing generic linear transformations via singular value decomposition (SVD), in which a matrix is decomposed into unitary and diagonal matrices \cite{tezak2019integrated,prabhu2020accelerating,zhu2024dynamically,wu2024programmable}. In practice, reconfigurable matrix operations are achieved using electro-optic, thermo-optic, or optomechanical actuation \cite{wang2023microring}. 

Hybrid photonic–electronic integration has recently emerged as a promising approach to improve scalability and reduce time-to-solution \cite{hua2025}. For example, a time-multiplexed architecture using cascaded TFLN modulators has been demonstrated as a pathway to mitigating scaling limitations (Fig.~\ref {fig:1}c) \cite{al2025programmable}. The large bandwidth, low V$_{\pi}$, and strong electro-optic nonlinearity of TFLN modulators enable stable, high-speed operation. More generally, integrating multiple functional components on a single chip can improve scalability and robustness while preserving compactness and cost-effectiveness \cite{gao2024photonic,hua2025}. Supplementary Section 1.3 provides further details on recent integrated photonic Ising architectures.

\begin{table}[!htbp]
    \centering
    \small
    \renewcommand{\arraystretch}{1.2}
\label{tab:my_label}
    \begin{tabular}{ccccccccc} \hline \hline
     \textbf{Platform} & \textbf{Scheme} & \textbf{Connection}&  \textbf{Re.  Coupling}&  \textbf{Scalability}&  \textbf{Latency}&  \textbf{Spins}& \textbf{Year [ref.]}\\ \hline 
   \multirow{8}{*}{Fiber optics }&Fiber loop&  All-to-all&  Yes &  Good &  1.6 $\mu$s&  100& 2016\cite{mcmahon2016fully}\\ 
  && All-to-all & Yes & Excellent& 5 $\mu$s & $2 \times 10^3$&2016\cite{inagaki2016coherent}\\ 
&& Sparse& No & Excellent & 5.2 $\mu$s& $>10^4$&2016\cite{inagaki2016large}\\
  && All-to-all & Yes& Excellent& 24. 7 $\mu$s& $10^5$&2021\cite{honjo2021100}\\
  && All-to-all & Yes& Good& 3.5 $\mu$s& $256$&2025\cite{jin2025kerr}\\
           &MZM&  All-to-all&  Yes&  Good& $\sim$ 1 s &  100& 2019\cite{bohm2019poor}\\ 
         &&  Sparse &  Yes&  Excellent&  NA &  $2.56 \times 10^4$& 2022\cite{cen2022large}\\
         && \textcolor{black}{ All-to-all} &  \textcolor{black}{Yes}& \textcolor{black}{ Good}&  \textcolor{black}{2.71 ms} &  \textcolor{black}{4096} & \textcolor{black}{2026\cite{hu2026programmable}}\\
  && All-to-all& Yes& Good& 0.43 ms& 64&2023\cite{mwamsojo2023optoelectronic}\\ \hline
   & Delay line &  All-to-all&  Partial &  Poor&  16 ns &  4& 2014\cite{marandi2014network} \\ 
      && Sparse &  No &  Poor &  110 $\mu$s&  16 & 2016\cite{takata201616}\\ 
       \multirow{10}{*}{Free-space optics }   &MMF&  Sparse&  No&  Poor&  NA &  13& 2019\cite{babaeian2019single}\\ 
           &SLM &  All-to-all&  Yes &  Excellent&  NA&  $7.5 \times 10^4$& 2019\cite{pierangeli2019large}\\ 
           & &  All-to-all&  Yes&  Good&  NA&  100& 2020\cite{pierangeli2020adiabatic}\\ 
           &&  All-to-all&  Yes &  Good&  NA &  100& 2020\cite{pierangeli2020noise}\\ 
      %      && All-to-all& No& Good & NA& 800&2020\cite{kumar2020large}\\ 
  && All-to-all& Yes& Excellent& NA& $ 4 \times 10^4 $&2021\cite{huang2021antiferromagnetic}\\
  && All-to-all& Yes& Good & NA& 100&2021\cite{fang2021experimental}\\
  && All-to-all& Yes & Poor & 320 ms& 30 &2022\cite{ouyang2022demand}\\
  && All-to-all& Yes & Excellent& 1s & 100 &2022\cite{zhang2021quadrature}\\
   && All-to-all & Yes & Poor & 6.4 $\mu$s &16& 2025\cite{kalinin2025analog}\\
   && All-to-all & Yes & Poor & 10 ms &16& 2023\cite{WDM_Li2023}\\
    && All-to-all& Yes & Excellent & 3.25 s& $2 \times 10^4$&2023 \cite{ye2023photonic}\\ 
    && \textcolor{black}{All-to-all}& \textcolor{black}{Yes} & \textcolor{black}{Excellent} & \textcolor{black}{0.62 s} & \textcolor{black}{$3.6 \times 10^5$}&\textcolor{black}{2025 \cite{yao2025enhanced}}\\ 
    &Interferometer& Sparse & Yes & Good & NA & 1024 &2025\cite{gao2025all} \\ \hline 
  \multirow{6}{*}{Integrated photonics} &MRR& Sparse& Yes& Poor& $<$ 310 ns& 2&2020\cite{okawachi2020demonstration}\\
  &MZI& All-to-all& Yes& Poor& 1-0.1 ns & 4&2020\cite{prabhu2020accelerating}\\
  && Sparse& Yes& Poor& 1.71 ns & 4& 2025\cite{wu2025monolithically}\\
    &MZM&Sparse& Yes& Excellent & 370 $\mu$s& $1.63 \times 10^4$ &2024\cite{li2024scalable}\\ 
    &&Sparse& Yes & Poor & 265.1 ns & 16 &2024\cite{Ouyang24}\\ 
     && Sparse & Yes & Good & 3 ns & 64 &2025\cite{hua2025}\\
    &&Sparse& Yes & Excellent & 1.9 ms & 4.1 $\times 10^4$ & 2025\cite{al2025programmable}\\ 
   \hline \hline 
\end{tabular}

\caption{Various photonic Ising architectures. Here, we present only the largest spin size or novel architectures and algorithms across different platforms. Re. Coupling = Reconfigurable coupling, MMF = Multi-mode fiber, SLM = Spatial light modulator, MRR = Microring resonator, MZI = Mach–Zehnder interferometer, MZM = Mach–Zehnder modulator, NA = Not available. Here, latency refers to the per-step computation time.}
\label{tab:1}

\end{table}

\begingroup
\fontsize{10}{12}\selectfont
\begin{tcolorbox}[graybox, float, floatplacement=htbp]
\textbf{Box 2 | Hardware challenges}
Photonic Ising machines have strong potential as heuristic solvers for computationally hard optimization problems. However, most current architectures face critical limitations in scalability, reconfigurability, time-to-solution, and solution accuracy. These limitations remain major obstacles to million-spin implementations. 

\begin{multicols}{2}
\textit{a. Scalability:}
Scalable photonic architectures are essential for realizing practical Ising machines, but scalability is limited by device integration complexity, optical-electrical interfacing bottlenecks, and fabrication constraints, all of which become increasingly important at larger scales~\cite{su2023scalability}. Fiber-based Ising machines employing measurement feedback and time-multiplexing have demonstrated reconfigurable all-to-all spin connectivity and currently support some of the largest photonic Ising machines, with up to 100,000 spins. Despite their scale, these systems suffer from phase fluctuations, stability limitations, and long convergence times~\cite{honjo2021100}. Optoelectronic architectures offer improved phase stability, but their speed is constrained by data conversion, memory access latency, and electronic bandwidth, which can impede further scaling~\cite{bohm2019poor}. 

Spatially multiplexed free-space architectures using SLMs have achieved more than 10,000 spins with all-to-all connectivity but are limited by slow SLM refresh rates and nonlinear phase responses. Integrated photonic Ising machines, owing to their compact footprint, high-speed operation, and potential for co-integration with electronics, provide a promising route towards scalable architectures. However, they remain limited by design trade-offs, fabrication constraints, and low-bit precision ~\cite{prabhu2020accelerating}. Overcoming hardware bottlenecks such as phase instability, slow device response, unwanted nonlinearities, large- scale integration complexity, and electronic constraints is therefore essential for scalable large-scale photonic Ising machines. Commercial ASICs, digital signal processing (DSP) pipelines, and analog electronics could help mitigate some of these challenges and improve system performance. Driven by practical applications, ongoing research is advancing toward million-spin photonic Ising machines through advances in device integration, architectural optimization, and system-level co-design.
\\

\textit{b. Reconfigurability:}
A general-purpose Ising machine~\cite{albertsson2023highly} must support reconfigurable coupling matrices to address a broad class of optimization problems. For example, the Boolean satisfiability problems can map naturally onto Ising networks with higher-order interactions, beyond pairwise spin couplings, motivating extensions of conventional architectures to higher-order Ising machines. However, implementing large, arbitrary coupling matrices with efficient feedback remains a major challenge and requires significant architectural innovation. Coupling matrices can be weighted (e.g., with random, bimodal, or Gaussian- distributed weights) or unweighted with uniform weights. Supporting both cases is important for adapting the machine to diverse problem types and scales~\cite{kalinin2018simulating}. Hardware advances, including TFLN-based SLMs and electro-optic modulators, have enabled highly reconfigurable photonic Ising machines capable of addressing a broad range of optimization tasks. For all-to-all connectivity, however, the coupling matrix size scales as $N^2$ for $N$ spins, creating major demands on memory, control, and interconnects. Each multiplexing strategy introduces trade-offs: wavelength multiplexing increases system complexity; time multiplexing introduces latency; and spatial multiplexing increases footprint. One possible mitigation is to transform all-to-all connectivity into sparse connectivity by introducing auxiliary spins, thereby reducing the number of direct neighbours per node while preserving the functional structure of the original coupling matrix~\cite{aadit2022}. However, sparsification requires additional spins, introduces computational overhead, and can degrade performance, particularly when finding the ground-state solution for large-scale problems. 
\\

\textit{c. Frustrated energy landscape:}
Large-scale combinatorial optimization problems often exhibit frustrated energy landscapes, characterized by numerous local extrema that hinder convergence to the global optimum. Whether implemented through dynamical evolution or annealing algorithms, Ising machines can become trapped in local minima, reducing the probability of finding the optimal solution. Multiple runs with randomized initial conditions are therefore commonly employed to increase the likelihood of reaching the global minimum. Post-processing techniques can also help escape local minima~\cite{pirnay2024self}. These include perturbations from external fields~\cite{bachschmid2016variational}, which nudge the system away from local traps; data augmentation~\cite{lim2025data}, which generates modified instances from the original dataset; and local search algorithms~\cite{cen2022tree}, which iteratively explore nearby configurations rather than the entire solution space. In photonic systems, controllable optical noise can be injected optically, for example, through optical amplifiers or electronically to broaden the search space. Optical noise can be energy-efficient, whereas electronic noise provides reconfigurable and tunable noise profiles~\cite{al2025programmable}. Performance can be further improved by combining complementary algorithmic strategies, such as noise-induced annealing, in which noise intensity is gradually reduced; momentum-based updates; and chaotic control, in which deterministic chaos is introduced to help explore the energy landscape~\cite{fan2023photonic,leleu2021,kalinin2025analog}. Another challenge in photonic Ising machines is the freezing effect, in which spin configurations become fixed and stop evolving~\cite{bohm2018}. This can be mitigated by dynamically tuning hyperparameters such as feedback strength, introducing tunable noise, and incorporating additional nonlinearity\cite{bohm2019poor,pierangeli2020noise,bohm2021order}.
\end{multicols}
\end{tcolorbox}
\endgroup

\begingroup
\fontsize{10}{12}\selectfont
\begin{tcolorbox}[graybox, float, floatplacement=htbp]
\textbf{Box 3 | Algorithmic challenges}
\begin{multicols}{2}

\textit{a. Computational complexity:} Success probability and time-to- solution (TTS) are key metrics for evaluating the ability of Ising machines to find optimal solutions ~\cite{kalinin2022}. Success probability is the likelihood of reaching the ground-state solution in a single run and is typically estimated from many randomized trials. It depends on the algorithm, the nature of the problem (e.g., sparse vs dense graphs, weighted vs unweighted), and problem size $N$. For many problem classes, especially those with all-to-all connectivity, the success probability can decrease rapidly with increasing problem size, often following an approximately exponential form, $P_s \propto A(e^{-bN^x})$, where $A$, $b$, and $x$ are problem- and system-specific constants. TTS commonly represents the expected time to reach the ground state with a target success probability, often 99$\%$. It is estimated as where 
\begin{equation*}   
\text{TTS} = T_{\text{an}} \frac{\log_{10}(0.01)}{\log_{10}(1 - P_s)}
\end{equation*}
where $T_{\text{an}}$ is the annealing time and $P_s$ is the single-run success probability. Hyperparameters such as feedback strength, noise level, and annealing schedule must be optimized to balance TTS and success probability. For small-scale problems, short annealing times can be sufficient because success rates are often relatively high. For larger and more complex problems, longer annealing times may be needed to explore the solution space and achieve a high success probability. 

Electronic integration provides flexibility and control, but it can also introduce latency that degrades TTS. Key contributors include memory-access delays, electronic communication overhead, and DSP latency. These overheads become increasingly significant at a large scale and can limit both scalability and overall performance. Recent strategies for reducing TTS include: implementing low-latency, parallel feedback systems~\cite{hua2025}; incorporating machine learning-enhanced annealing strategies~\cite{cossu2019machine}; and employing diverse nonlinear transfer functions, which have been shown to accelerate convergence ~\cite{bohm2021order}. 
\\
\textit{b. High-quality solutions:}\\
Ising machines iteratively evolve towards low-energy configurations, with solution quality ideally improving over successive iterations. However, low bit precision, amplitude heterogeneity, signal degradation, synchronization errors, and phase instability can degrade the quality of the final solution. In addition, the exponential scaling of TTS with problem size can reduce practical accuracy by limiting the number of feasible runs or iterations. 

Depending on the problem, an Ising machine may have multiple degenerate ground-state solutions. Identifying these configurations can be valuable in applications such as materials design, drug discovery, and machine learning, where multiple valid solutions can provide additional insights~\cite{weinand2022research}. For problems with degenerate ground states or many near-optimal solutions, sampling techniques, in which the machine is run repeatedly under slightly varied conditions, can help uncover multiple valid configurations. Weighted-sum objective functions can also combine competing objectives into a single scalar objective using predefined weights, enabling exploration of trade-offs among different solution criteria~\cite{pal2020rapid,ng2022efficient}.

High bit precision is generally desirable for accurately mapping a target problem onto the Ising Hamiltonian. Nevertheless, photonic Ising machines with modest precision can sometimes achieve performance comparable to, or better than, digital electronic solvers operating at higher floating-point precision, depending on the problem, dynamics, and noise properties ~\cite{sevenants2025requirements}. Ising solvers also involve many hyperparameters, and their performance can be sensitive to these choices. Identifying optimal hyperparameter settings is often time-consuming and computationally expensive. Dynamically adjusting hyperparameters during optimization can reduce computational time while maintaining high-quality approximate solutions, improving the practicality of Ising machines for real-world applications. 

\end{multicols}
\end{tcolorbox}
\endgroup

\pagebreak

\section{Roadmap to a million-node Ising processor and beyond}
\label{sec:million}
Despite sustained progress in computing hardware and algorithms for complex optimization problems, it still requires high-speed computation, large-scale memory access and resource-intensive models ~\cite{dara2022, ofuchi2021}. These requirements limit scalability, computational efficiency and can prolong convergence to high-quality solutions for large-scale, real-world optimization tasks. Drug discovery provides a representative example. Conventional drug development typically spans 12–15 years due to iterative trial-and-error workflows and involves substantial research and development costs~\cite{hughes2011}. Mapping aspects of drug discovery to the Ising model offers an alternative route, for example by identifying candidate compounds through the combination of functional fragments as low-energy, ideally ground-state solutions ~\cite{xiong2023q,torquato2011toward,liang2025p}. Such formulations could improve computational speed and solution quality, but they require millions of Ising spins. 

Photonic architectures offer low latency, parallelism, and energy efficiency, but realizing a reconfigurable million-node photonic Ising machine remains a major challenge. Most proposed platforms require advances in dense photonic integration, packaging, control electronics, and algorithms. Several routes are being explored, including all-optical or optoelectronic systems based on spatial, temporal, or hybrid multiplexing. Emerging approaches such as integrated photonic chiplets, all-optical analog Ising architectures, and coherent spatiotemporal Ising processors offer promising pathways toward next-generation scalable photonic Ising machines. Figure~\ref{fig:1}(d) summarizes a possible roadmap towards million-node Ising machines and quantitatively compares different Ising processor architectures in terms of number of spins, bit precision, frequency, and latency. The accompanying spider graph contrasts CMOS-based systems with proposed photonic Ising machines across qualitative design metrics, including scalability, reliability, speed, reconfigurability, and footprint.
\subsection{Photonic chiplet Ising processor}
\label{sec:roadmap}
\begin{figure}[htbp]
    \centering
        \includegraphics[height=6.8in]{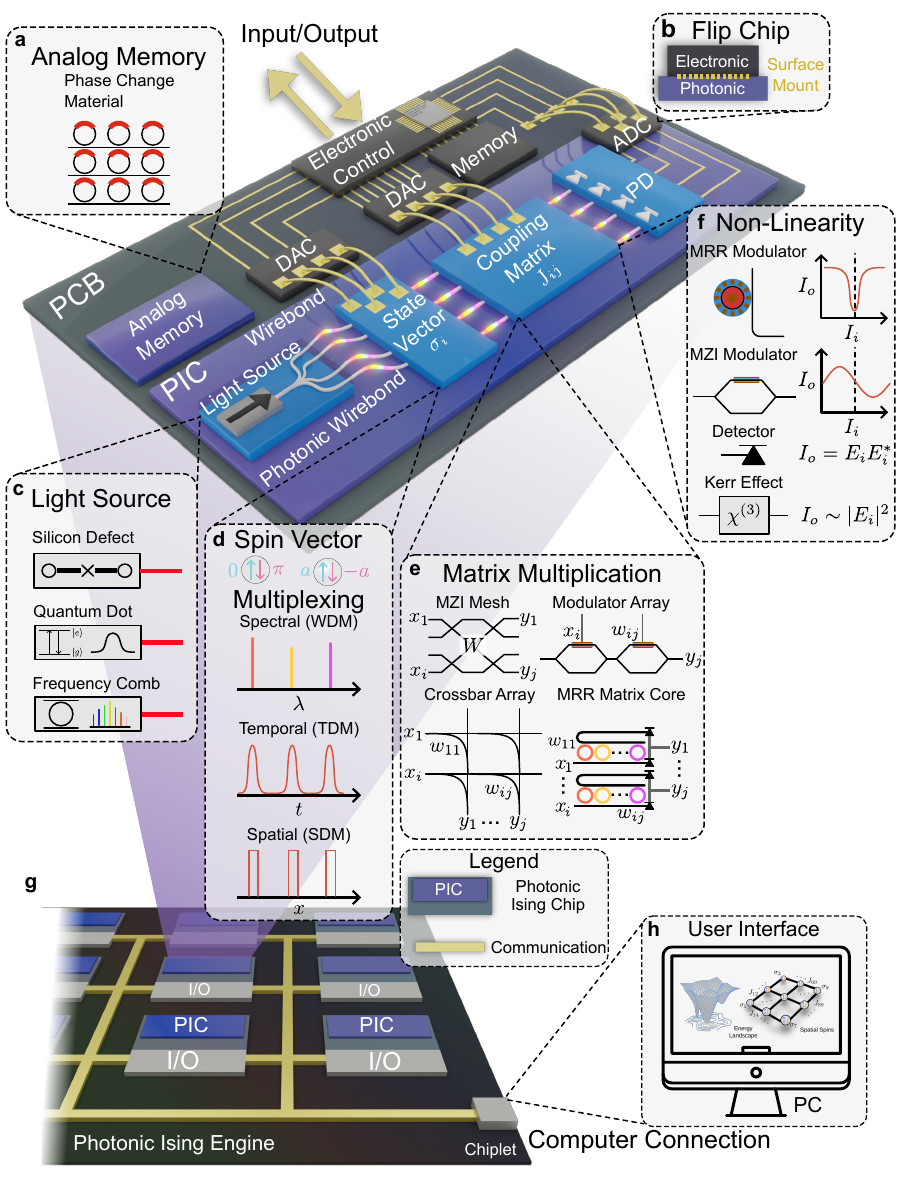}
    \caption{\textbf{ Photonic chiplet Ising architecture:}
    A conceptual on-chip processor utilizing commercially available photonic technologies. The programmable processor integrates on-chip photonic components, co-integration, packaging, and I/O strategies. \textbf{a)}  Optical non-volatile analog memory made of phase-change materials. \textbf{b)} Flip-chip bonded ASIC on photonic integrated circuit for wirebond-free electronic control. The ASIC configures the ADCs and DACs to enable high-speed detection, digital signal processing, and driving of the electro-optic components. Gold lines represent the electrical connections. \textcolor{black}{\textbf{c)} Different solutions for the proposed on-chip light sources include frequency combs, III-V lasers, and quantum dots. \textbf{d)} Ising spin vectors can be realized by utilizing different degrees of freedom, such as spectral, temporal, or spatial modes.} \textbf{e)} Matrix multiplication using various on-chip architectures, including Mach-Zehnder interferometers, meshes, modulator arrays, crossbar arrays, or microring modulator matrix cores, enabling  low-latency and high-speed MVMs. \textbf{f)}  Realization of on-chip optical nonlinearity using well-known components such as microring resonators, Mach-Zehnder interferometers, or intensity detectors. \textbf{g)} The chiplet comprises several sub-processors. The electronic control unit interfaces with various components, such as a microcontroller, co-located digital memory, and DACs. \textbf{h)} User interface to control the Photonic Ising Engine.}
    \label{fig:On-Chip}
\end{figure}

Progress in integrated photonics has enabled photonic systems with higher throughput, lower energy consumption, and smaller physical footprints. Recent demonstrations based on silicon and TFLN photonics have shown promising progress toward scalable integrated photonic Ising machines \cite{hua2025,al2025programmable}. However, current hardware remains far from achieving a million-spin system. Realizing large-scale all-to-all spin connectivity on a single photonic chip remains technically challenging and costly because of component count, routing complexity, control overhead and footprint. 

Inspired by electronic chiplets ~\cite{sharma2022increasing, kashimata2024efficient,lan2024integrated}, photonic chiplet architectures offer a modular route to scalable photonic computing for real-world optimization tasks. In this approach, the interaction matrix $J$ of the Ising problem is partitioned across multiple chiplets by decomposing it into inter- and intra-chiplet interactions ~\cite{sharma2022increasing}. For example, an $N$-node problem can be divided into $ \ x\times y$ sub-arrays, with each sub-problem solved in parallel while preserving the original coupling matrix $J$ and local fields $h$ after decomposition ~\cite{xu2024large,tatsumura2021scaling}. Multiple photonic chips can therefore operate collaboratively, with each chiplet processing a submatrix of the overall Ising model~\cite {kashimata2024efficient}. \textcolor{black}{However, this decomposition introduces additional computational and communication overhead. For example, divide-and-conquer methods can decompose large-scale MVMs into smaller subproblems but require additional polynomial time of O$(n^{2.81})$ \cite{pan2013divide,izadkhah2022divide}}.

Each chiplet contains a subset of spins and the local logic needed to implement couplings among them. Although each chiplet performs local MVM in parallel, chiplets must frequently exchange updated partial sums and spin states to capture inter-chiplet couplings and suppress error accumulation. High-speed inter-chiplet communication is therefore essential. In principle, an $x\times y$ chiplet configuration, with each chiplet containing \textit{N} spins, could implement an \textit{xyN} spin Ising machine. A key advantage of this decomposition is that the full system can still be analyzed as a coupled dynamical system ~\cite{sharma2022increasing,yang2024sophie}. Although photonic chiplets can operate in parallel with nanosecond-scale latency, the overall system latency is also determined by electronics, memory access, and inter-chiplet communication. These bottlenecks are constrained by Amdahl’s law and increase the effective latency from nanoseconds to the microsecond regime ~\cite{hua2025,xu2024large}. The proposed on-chip photonic chiplet architecture, illustrated in Fig.~\ref{fig:On-Chip}, consists of an $x\times y$ array of chiplets. Each chiplet comprises optical interconnects, data converters, electronic control circuitry, optoelectronic components, and communication units to execute its designated segment of the Ising algorithm. 

Dense integration of optical and electronic components is essential for energy-efficient chiplet-based Ising processors. Co-integration with CMOS electronics and heterogeneous integration of photonic components provide practical approaches for programming, control, and readout. Wire bonding, flip-chip bonding, 2.5D integration, and 3D heterogeneous integration can enable very-large-scale integration and support the realization of large-scale systems~\cite{shastri2021photonics}. As shown in Fig.~\ref{fig:On-Chip}, a large-scale, on-chip photonic Ising processor would require integrated light sources, spin-spin coupling units, nonlinearities, detectors, and electronic control circuitry. For example, a $5\times 5$ chiplet architecture in which each chiplet handles >40,000 sparsely connected spins could collectively realize a million-node ($xyN$ spins) Ising machine, albeit with additional spin, communication, and computational overhead. However, an 18- inch wafer could accommodate monolithic integration of 64×64 multicore cascaded MZI modulators (each occupying $\sim$ 12 $mm^{2}$), potentially enabling an all-to-all-connected photonic Ising machine with more than one million nodes\cite{shi2023800g}. 

Further advances in multiplexing techniques, modular design, 3D-stacked architectures, and advanced packaging could reduce integration complexity while improving scalability and energy efficiency~\cite{xiang2024building,zheng2000free,ogura2023spatial}. Light sources, on-chip coupling matrices, and recurrent feedback form the core subsystems required for practical chiplet-based photonic Ising machines.

\subsubsection*{Light source}
Chiplet-based Ising architectures require independent on-chip light sources to drive each chiplet. Spatial, temporal, wavelength, and hybrid multiplexing can then be employed to encode spins in optical intensity or phase, enabling different subproblems to be assigned to individual chiplets. Recent advances in microring-resonator frequency combs, silicon defect centres, heterogeneous and monolithic integration of III-V materials, and quantum-dot-based on-chip emitters offer promising routes to integrated light sources for chiplet architectures (Fig. ~\ref{fig:On-Chip}c)~\cite{zhou2015chip, zhou2023prospects}. Among these technologies, integrated frequency combs are particularly attractive because a single comb source can provide more than 100 coherent wavelength channels aligned within the telecom ITU grid for highly parallel processing ~\cite{chen2015, al2025programmable,kovaios2025chip}. Broadband combs are typically generated in microring resonators through Kerr nonlinearity, requiring high-Q resonators, dispersion engineering, and low-loss material platforms, such as silicon nitride, hydex, or III-IV semiconductors ~\cite{kippenberg2011}. Such frequency combs have been commonly demonstrated in telecommunications, metrology, spectroscopy, and neuromorphic computing~\cite{gaeta2019, chang2022}. 

Compact on-chip III-V lasers provide complementary advantages, including high optical gain, thermal and operational stability, and compatibility with electrical pumping. Although III-V materials are not naturally CMOS-compatible due to lattice mismatch and differences in thermal expansion with silicon ~\cite{porter2023}, several integration and co-packaging strategies have been developed. These include flip-chip bonding, transfer printing, direct bonding, epitaxial growth of III-V materials on silicon, and hybrid integration techniques~\cite{billah2018, song2016, guan2018, wang2018, zhang2018}. Another promising direction is the direct integration of tunable quantum dot-light sources on silicon~\cite{liu2015}. Their nanoscale dimensions and quantum confinement effects make quantum dots well-suited for compact and tunable laser sources.

\subsubsection*{On-chip spin-spin coupling}
In chiplet-based Ising architectures, each chiplet requires an independent, reconfigurable spin-spin coupling unit to encode the coupling matrix. By leveraging parallel processing and hybrid photonic–electronic approaches, optoelectronic platforms have demonstrated computational throughput for MVM operations in the range of tera- to peta-operations per second~\cite{feldmann2021}. Reconfigurable photonic MVM cores implement unitary or non-unitary, real- or complex-valued matrix multiplication and can enable programmable spin-spin coupling across a broad range of computational problems~\cite{cheng2022small}. Integrated photonic MVMs can offer a compact footprint, broad bandwidth, reconfigurability, reduced system complexity, and high stability, making them attractive for high-performance computing \cite{zhou2022photonic,nahmias2019}. As a result, several integrated photonic architectures can support spin-spin coupling through spatial or temporal multiplexing. Figure~\ref{fig:On-Chip}e illustrates a representative chiplet with integrated photonic components for MVM, including MZI mesh arrays, photonic crossbar arrays, on-chip modulators, and microring resonator arrays. 

MZI meshes and photonic crossbar arrays use spatial multiplexing to implement compact and energy-efficient MVMs, although current demonstrations remain relatively small-scale. By contrast, optoelectronic architectures based on high-speed modulators can implement reconfigurable coupling matrices through time multiplexing. Typically, these modulators are based on pn-junction devices and silicon–germanium electro-absorption modulators fabricated in standard silicon foundries, as well as Pockels-effect devices based on lithium niobate (LiNbO$_3$), organic polymers, and barium titanate (BaTiO$_3$), offering operating bandwidths exceeding 100 GHz~\cite{talkhooncheh2022}. In particular, TFLN modulators have recently demonstrated modulation speeds above 110 GHz, strong thermal stability, and low insertion loss, highlighting their potential for chiplet-based Ising machines~\cite{mercante2016, al2025programmable}. Because these optoelectronic architectures perform MVMs across both optical and electronic domains, electrical co-integration is essential for control, feedback and data movement. Photonic MVMs based on incoherent or partially coherent light sources, implemented with crossbar arrays and microring resonator arrays, provide an alternative route. These systems can offer interference-free operation, high parallelism, low latency, and improved stability without substantially sacrificing accuracy and could support fast and scalable MVMs~\cite{dong2024partial,tait2017}.

\subsubsection*{Recurrent feedback} 
\textcolor{black}{Recurrent feedback enables dynamic spin updates after each matrix multiplication. Through successive iterations, the feedback loop drives the system through the Ising energy landscape towards low-energy and, ideally, global-state spin configurations. In photonic Ising machines, recurrence is typically implemented using either optical feedback or optical-electronic-optical (O-E-O) feedback\cite{mcmahon2016fully,bohm2019poor}. These feedback mechanisms also provide a route to integrate reconfigurable nonlinearities and DSP, both of which can improve convergence, solution quality and algorithmic flexibility.} 

\textcolor{black}{\textit{Nonlinearity:} The choice of a nonlinear function strongly influences solution accuracy, convergence dynamics, and the ability of the system to escape local minima. Nonlinear functions such as sinusoidal, polynomial, and sigmoidal activations can be implemented digitally using FPGAs or ASICs at each iteration. For example, using a sigmoidal nonlinearity rather than a polynomial one has been shown to improve TTS by an order of magnitude~\cite{bohm2021order}. Optical or optoelectronic nonlinearities can also be embedded directly into the dynamical system, enabling ultrafast analog signal processing without digital conversion. Figure~\ref{fig:On-Chip}f illustrates commonly used nonlinearities in photonic Ising machines, including electro-optic nonlinearities (e.g., sinusoidal, sigmoidal, or square activation functions) and second- and third-order optical nonlinearities~\cite{tait2019, williamson2019, mourgias2019, bohm2021order}.} 

\textcolor{black}{\textit{Digital signal processing:} DSP is critical for improving the performance of large-scale, high-speed photonic Ising machines. In time-domain multiplexed architectures, latency scales directly with the spin duration; therefore, increasing the signaling rates can proportionally reduce latency and TTS. In this context, the signaling rate is analogous to the symbol rate in optical communication systems. However, higher symbol rates impose stringent bandwidth requirements on analog electronic components, including DACs, ADCs, RF amplifiers, photodiodes, and RF packaging.} \textcolor{black}{System-level impairments caused by bandwidth limitations and undesired nonlinearities, such as RF driver compression, can degrade overall system performance\cite{DiCheDSP}.} \textcolor{black}{These impairments are particularly important because they occur at solver iteration and can therefore accumulate over time. For example, an uncompensated system-wide frequency response filters the time-domain spin vector at each multiplication iteration. For the linear components of the system response, an uncorrected frequency response $H(f)$ compounds over N iterations as $H(f)^{N}$. This accumulation creates a need for signal-processing algorithms that both optimize the spectral shape of the time-domain spin-vector representation and compensate for compounding distortions.} 

\textcolor{black}{DSP algorithms developed for optical communication systems can be adapted to address these impairments. At the transmitter, pulse-shaping the signal with Nyquist pulses reduces the required bandwidth to as low as half the symbol rate. Bandwidth limitations can be corrected adaptively using a short pilot sequence and a feed-forward equalizer (FFE), whereas system nonlinearities can be compensated using polynomial nonlinear equalizers. In time-multiplexed photonic Ising machines, such DSP techniques can increase symbol rates by orders of magnitude, thereby reducing TTS. Pulse shaping combined with FFE has also been shown to improve solution quality at higher symbol rates by compensating for system impairments\cite{al2025programmable}. In addition, controlled residual inter-symbol interference from non-ideal pulse shaping and digital filtering can induce neighboring-spin interactions, which may aid computation\cite{al2025programmable}. However, DSP also increases ASIC footprint, adds latency to the feedback path, and raises digital power consumption.} \textcolor{black}{Photonic chiplet architectures face several practical challenges. Unlike modern CMOS chiplet systems, in which multicore compute blocks can be monolithically integrated with high manufacturing yield, photonic systems typically require co-integration of multiple material platforms, complex device design, and precise optical packaging. Realizing million-node photonic Ising machines will therefore require dense integration of photonic components, scalable packaging technologies, advances in algorithms, and large-scale optical and electronic control systems.}

\subsubsection*{Practical challenges in chiplet architecture:}
\textcolor{black}{Unlike modern CMOS chiplet architectures, where entire multicore compute blocks can be monolithically integrated, photonic systems rely on the cointegration of different material platforms, require complex device design, and demand high manufacturing yield. In addition, the need for dense photonic component integration, packaging technologies, advances in algorithms, and the development of large-scale optical and electronic control units remains a major challenge. Similarly, careful design of high-bandwidth photonic devices, mitigation of fabrication imperfections, and compensation for device losses are required to avoid any error accumulation. As the problem is partitioned into sub-problems, each chiplet contains a subset of spins associated with a given sub-problem. As a result, the system requires additional overhead in terms of the number of spins and computational resources for sub-problem mapping. To efficiently handle large amounts of data, each chip requires a distributed memory unit to store partial results and coupling weights. In addition, low-loss optical interconnects, data converters, electronic control units, high-speed inter-chip communication and precise synchronization across chips are crucial for its performance~\cite{yamamoto2022scalable,xu2024large,takemoto20214}.}

\subsection{Free-space photonic Ising processor}
\label{sec:million}
\begin{figure}[ht!]
    \centering
    \includegraphics[]{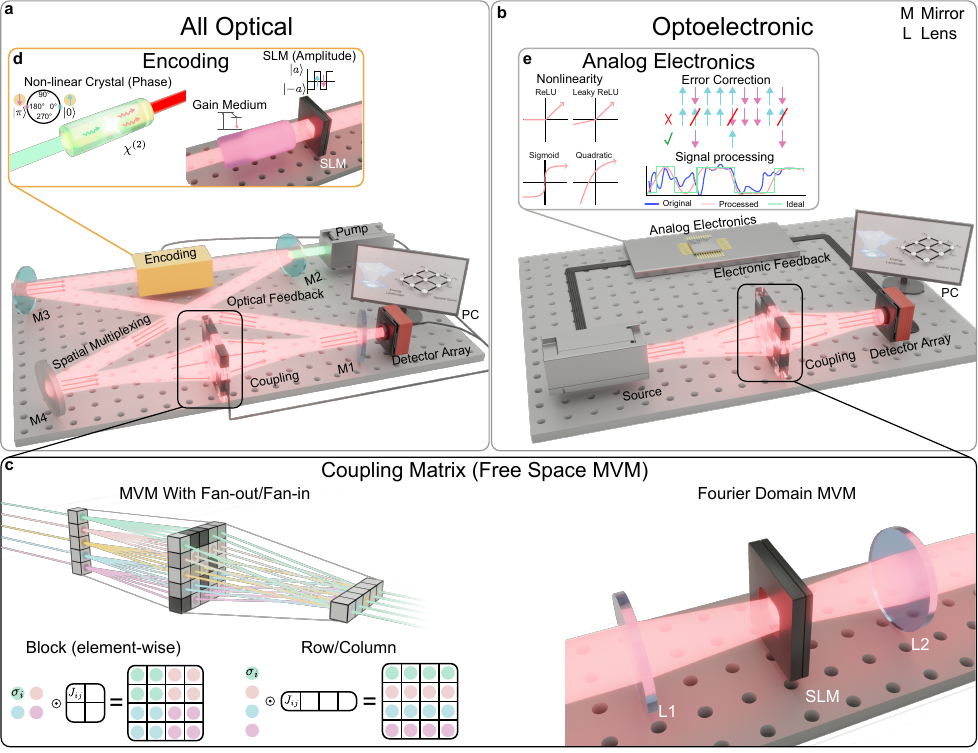}
    \caption{\textbf{Spatially multiplexed photonic Ising architecture:} a) An all-optical Ising machine architecture implemented using optical spin encoding, optical MVM, and optical feedback. b) Optical spin encoding realized through either degenerate OPOs or laser encoding, where the gain-based system naturally converges to minimum-energy spin configurations. c) Large-scale all-optical free-space MVM achieved using spatial multiplexing schemes, implemented through fan-out/fan-in or Fourier-domain multiplexing. d) An alternative, hybrid optoelectronic Ising machine designed with optical encoding, spatially multiplexed MVM using free-space components, and optoelectronic feedback. e) The optoelectronic feedback allows either analog or digital electronics to facilitate the incorporation of nonlinearity, signal processing, and the implementation of novel algorithms.}
    \label{fig:6}
\end{figure}

Spin encoding and updating, MVM, and nonlinearity are among the most time- and energy-intensive operations in Ising machines when they are not implemented optically  ~\cite{bohm2019poor,honjo2021100}. At large scale, time multiplexing typically increases latency due to sequential processing and I/O overhead in the digital domain. By contrast, free-space spatial multiplexing can perform large MVMs in parallel and offers low latency. Free-space MVM is among the earliest approaches in optical computing and has been used in optical logic processors, shadowgram, and other optical computing platforms~\cite{goodman1978fully,tanida1983optical}. In spatial multiplexed Ising architectures, spin generation, matrix multiplication, and feedback can, in principle, be carried out entirely in the optical domain or with minimal optical–electronic conversion. This makes free-space optics attractive for scalable, high-speed, and energy-efficient Ising processors. \textcolor{black}{Here, we focus on two related approaches: all-optical Ising processors (Fig. ~\ref{fig:6}a) and optoelectronic Ising processors (Fig. ~\ref{fig:6}b). Both leverage the inherent parallelism of free-space optics for scalable MVM. However, these architectures remain conceptually limited or have been demonstrated only at small scales, for example, an optoelectronic Ising machine with 16 spins}. 

Similar to other architectures, an $N$-spin system requires $N^2$ matrix elements to implement an arbitrary coupling matrix in the free-space implementation. However, as shown in Fig. ~\ref{fig:6}c, the optical fan-in and fan-out requirements redundantly encode the spatial input across $M_x$ pixels, perform element-wise multiplication, and then sum row-wise over $M_y$ to realize the desired MVM~\cite{spall2020fully}. This method enables reconfigurable, arbitrary connectivity but limits the number of encoded spins to $N = M_x$. An alternative is to perform the computation in momentum or Fourier space, avoiding explicit fan-in and fan-out when the matrix-vector operation can be expressed as a convolution ~\cite{ calvanese2021all,davis1994circulant}. As shown in Fig. ~\ref{fig:6}c, a transfer-matrix mask placed at the focal plane of the first lens $L_1$ performs computation in momentum space via a Fourier transform, enabling element-wise multiplication of the spatial-frequency components of the incident oscillator amplitudes. A second lens, $L_2$, identical in focal length to $L_1$, performs an inverse Fourier transform, returning the signal to real space. According to the convolution theorem, multiplication by the transfer function in the Fourier domain implements a convolution in real space, thereby realizing the coupling operation defined by the mask~\cite{calvanese2021all}. In this configuration, the number of encoded spins can scale as $N = M_x \times M_y$, where $M_x$ and $M_y$ are the numbers of spatial modes along the two transverse directions. However, this approach is naturally suited only to problem classes whose spin interactions can be expressed as convolutions, such as nearest neighbour Ising chains and Möbius ladders. 

In general, SLMs can provide programmable weight matrices, but their scalability is limited by a finite pixel count (typically on the order of $10^6$). In contrast, metasurfaces and other diffractive optical elements, composed of densely patterned micro- or nanostructures, could support millions of Ising spins\cite{ luo2026highly, park2024all, stanley2003100}. Additionally, integrating non-volatile materials, such as phase-change materials (PCMs) or photochromic compounds, could further enable energy-efficient, reconfigurable coupling matrices\cite{zhang2026non,xu2025building}. Further, the third spatial dimension (z-axis) enables efficient all-optical fan-in and fan-out, potentially facilitating the implementation of large-scale Ising solvers across all-optical and optoelectronic free-space platforms.

\subsubsection*{All-optical Ising processor:}
Degenerate optical parametric oscillators (DOPOs) and injection-locked coupled laser systems are promising all-optical platforms for emulating the Ising model because their dynamics naturally evolve towards low-energy solutions~\cite{ising1924beitrag}. In these systems, gain, loss, coupling, and nonlinear saturation collectively determine the steady-state configuration. Near the oscillation threshold, configurations with the lower effective loss corresponding to the lower Ising energy are preferentially amplified, providing a physical mechanism for heuristic optimization (see Fig. ~\ref{fig:6}a). However, existing all-optical coherent DOPO architectures support only a limited number of spins because of limitations associated with time-multiplexing, stability and latency, which limit scalability and their applications to real-world problems~\cite{marandi2014network,mcmahon2016fully,pal2020rapid}. Incorporating spatial MVM into these systems potentially overcomes these limitations and supports a reconfigurable interaction matrix $J_{ij}$. Recently proposed spatially multiplexed all-optical parametric oscillators or spatially encoded CIMs could reduce electronic overhead by enabling simultaneous encoding and updating of all spins \cite{calvanese2021all}. In these systems, spatially multiplexed degenerate parametric oscillators based on second-order nonlinear crystals ($\chi^{2}$), encode bistable spin states in the phase front of the optical beam, typically as phases of 0 and $\pi$~\cite{calvanese2021all} (see Fig. \ref{fig:6}d). An array of $N$ spatial parametric oscillators can therefore encode many independently controllable spins across the transverse XY plane. 

An alternative all-optical approach uses a coupled-laser system with spatially multiplexed spin encoding in the laser beam profile \cite{nixon2013observing,pal2020rapid}. In these architectures, spins are represented by the phase or intensity profile of an array of spatial lasers, forming a network of spatially encoded spins (see Fig. \ref{fig:6}d) \cite{utsunomiya2011mapping,takata2014data}. \textcolor{black}{In principle, millions of spin configurations could be realized in spatially multiplexed coupled laser systems, with the coupling matrix implemented via controlled interactions among laser modes within the optical cavity \cite{nixon2013observing} (see Figure~\ref{fig:6}c). The resulting driven–dissipative dynamics iteratively evolve the spin configuration, with internal dissipation guiding the laser network towards stable spatial intensity patterns that correspond to low-energy configurations of the Ising Hamiltonian. The main challenge is implementing a large-scale, reconfigurable, and accurate coupling matrix. Possible strategies include adaptive optics, SLM, and additional correction matrices. These architectures can be implemented using conventional laser systems or VCSELs. For a representative cavity length of $\sim$30 cm, the round-trip rate can reach the GHz regime, corresponding to nanosecond-scale iteration times and an overall TTS in the microsecond regime. For example, a 1,000-spin vector with one million coupling elements in a 1 GHz optical cavity performs approximately $10^{6}$ MAC operations per cavity round trip. The effective computational throughput can therefore approach $\sim$2 PFLOPS, comparable to the FP8 Tensor Core performance of the NVIDIA H100, which is approximately $\sim$1.98 PFLOPS\cite{nvidiaNVIDIAH100Tensor}. However, the estimated power consumption of the photonic system is only $\sim$13 W, including the power required to store the matrix elements in the SLM and the injected optical power, compared with $\sim$700 W for the NVIDIA H100  GPU\cite{nvidiaNVIDIAH100Tensor}.}

\subsubsection*{Optoelectronics Ising processor:} 
Figure ~\ref{fig:6} (d) shows an optoelectronic Ising architecture in which the computationally intensive large-scale MVM is performed in the optical domain via spatial optical modulation, whereas nonlinearity and feedback are implemented electronically. This division of functions combines the parallelism of spatially multiplexed optical MVM with the flexibility of electronic feedback, offering advantages in spin encoding, connectivity, nonlinear response, and algorithmic design. As a result, optoelectronic Ising processors can be adapted to a broader range of optimization tasks than fully passive optical systems. A fully analog optoelectronic architecture, combining analog optical and analog electronic circuits, offers the additional advantage of avoiding data transfer, memory access, and repeated conversion between digital and optical domains, thereby minimizing latency and improving scalability~\cite{solli2015analog,mohseni2022ising,zhu2023intelligent}. Although O-E-O conversion is still required, it can be implemented with standard, off-the-shelf discrete optoelectronic components and is typically less expensive than high-speed digital conversion. \textcolor{black}{A recent analog optical computer integrates analog electronics with three-dimensional optics to accelerate combinatorial optimization through rapid fixed-point search ~\cite{kalinin2025analog}}. In the analog optoelectronic Ising architecture, spin vectors can be encoded in the spatial intensity profile of inexpensive light sources, such as lasers, LEDs, or incoherent emitters driven by electrical signals~\cite{wang2022optical}, (see Figure~\ref{fig:6}d). The coupling matrix can be implemented using a spatially multiplexed scheme similar to that shown in Fig. ~\ref{fig:6}c, where input spin vectors are encoded in the intensity profile of the light beam, matrix elements are encoded in the transmission profile of the spatial $N \times N$ array, and an optical fan-in and fan-out replicate the \textit{N} spin variables\cite{kalinin2025analog}. In simple intensity encoding, the coupling matrix is restricted to nonnegative weights, thereby limiting the class of directly solvable problems~\cite{chen2023all}. However, offsetting and scaling procedures can be used to represent arbitrary signed matrix weights in the analog optical domain. As shown in Fig. ~\ref{fig:6}e, analog electronic feedback further enables user-defined nonlinearities, advanced iterative algorithms, and analog signal processing without requiring conversion to digital signals.     

In analog optical–electronic co-integrated architectures, latency is primarily determined by the optical path length and operational bandwidth, thereby supporting processing speeds in the GHz range. \textcolor{black}{A recent optoelectronic demonstration employed a 16-LED array and an SLM to implement a 16-spin processor for small-scale optimization problems. Scaling beyond this regime will require larger spatial modulators, detector arrays, and emitter arrays. Commercially available SLMs with more than ten million pixels, such as GAEA-2 devices with 4160$\times$2464 pixels \cite{Holoeye}, million-pixel CMOS detector arrays, and 2D micro-LED arrays with more than 1000 elements, could implement 1000s of spins, corresponding to one million coupling elements. By contrast, dense 2D micro-LED or OLED display technologies used in VR/AR, with pixel sizes approaching 1 $\mu$m and resolutions of 4800$\times$3840 pixels\cite{liu2020micro}, could in principle support a million-node (all-to-all coupling) Ising machine in Fourier space. However, such Fourier-domain implementations are restricted to coupling matrices that can be represented by convolutional operations. However, similar to chiplet architectures, multiple free-space modules can be implemented to partition large problems into subproblems and potentially allow the mapping of one million nodes. Nevertheless, mapping overhead and distributed coordination remain significant challenges}.

\subsubsection*{Practical challenges in free-space architectures:}
\textcolor{black}{ Several practical challenges remain for free-space photonic Ising machine architectures. Because spins are encoded in the spatial beam profile, these architectures require high beam quality and preferably flat-top intensity profiles to suppress spatial amplitude inhomogeneity, i.e., spatially varying spin amplitudes. Diffraction effects arising from non-uniform optical elements and alignment errors can introduce undesired spin-spin interactions, although these effects can be mitigated using optical correction matrices \cite{carstensen2026photonic}. The system must also maintain sufficient signal-to-noise ratio at readout to distinguish spin states reliably\cite{babaeian2019single}. More generally, a high signal-to-noise ratio and large dynamic range are essential for solution accuracy. Additional limitations include bulky optical setups and device-level imperfections such as pixel nonuniformity, phase quantization and SLM thermal drift. Addressing these issues will be essential for translating free-space optoelectronic Ising processors from small-scale demonstrations to scalable optimization hardware.}

\subsection{All-optical coherent spatiotemporal architecture:}
\begin{figure}[ht!]
    \centering
    \includegraphics[height=4.2in]{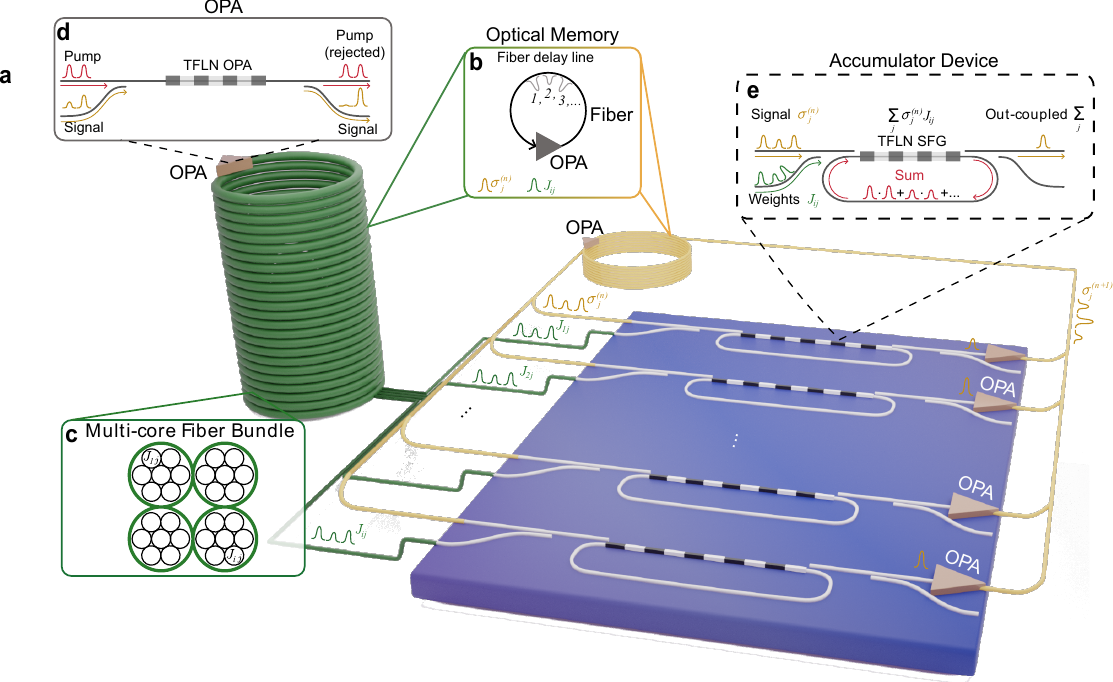}
    \caption{\textbf{All-optical coherent spatiotemporal architecture:} \textbf{a)} Schematic of the All-optical coherent spatiotemporal architecture implemented using time- and spatial-multiplexed configurations. \textbf{b)} Fiber-based optical memory used for storing large spin and coupling matrices. \textbf{c)} Shows the typical multicore fiber employed for storing the interaction matrix. \textbf{d)} Implementation of OPA to compensate for optical losses. The TFLN crystal is periodically poled to phasematch the 780 nm pump and 1560 nm signal wavelengths for the OPA process. \textbf{e)} Schematic of the accumulator cavity using a TFLN-based SFG process. The TFLN crystal is phasematched at 1560 nm (spin vector) and 1300 nm (weight) to generate a sum-frequency wavelength of 710 nm.} 
    \label{fig:Comparison}
\end{figure}

The spatiotemporal architecture exploits spatial and temporal degrees of freedom to implement a million-scale Ising machine. In the proposed all-optical coherent spatiotemporal approach (see Fig.~\ref{fig:Comparison}(a)), the spins are encoded directly in the field amplitudes of optical pulses throughout the computation, and the entire system dynamics, including memory access, arithmetic, and nonlinear functions, is realized all-optically via waveguided photonic circuits. As a result, neither optical-electronic-optical nor analog-digital-analog conversions are necessary in principle. For achieving large scalability, this architecture utilizes one temporal dimension, slicing up the loop iteration time into $O(N)$ time slots supporting $O(N)$ pulses, and one spatial dimension—consisting of \textit{N} waveguided channels, thus enabling all $O(N^2)$ operations in one loop iteration. Despite the increased latency, the ability to “coil up” the temporal dimension using ultra-low-loss fiber is a potentially useful way to compactify the overall spatiotemporal requirements of in-memory compute, particularly for $10^{12}$ Ising matrix elements. 

As shown in  Fig.~\ref{fig:Comparison}(a), the architecture consists of two parts, an optical memory unit implemented using fiber loop system and an optical processor unit realized with a TFLN integrated circuit~\cite{zhu2021integrated, boes2023lithium} to perform iterative computation. The Ising matrix elements are stored in a multicore fiber~\cite{mukasa2019100,chen2024applications} loop consisting of \textit{N} total fiber cores, with each fiber core storing \textit{N} pulses inside the loop (modulo some auxiliary header or control pulses) (see Fig.~\ref{fig:Comparison}(b)). Thus, there are a total of $O(N^2)$  different pulses circulating in this part of the memory, and each fiber core can be used to represent one column of the Ising matrix, while the elements of the column are temporally encoded as pulses in its respective fiber core (see Fig.~\ref{fig:Comparison}(c)). To overcome any optical loss in the fiber loop, a degenerate optical parametric amplifier (DOPA, see Fig.~\ref{fig:Comparison}(d)) can be used in-loop to balance the loss; the eventual SNR degradation from this process and technical phase noise in the mechanical stability of the memory ultimately limit the memory refresh time. Finally, readout from the fiber loop memory is achieved by outcoupling a fraction of the circulating power during each round trip of the loop from each fiber core; if necessary, additional DOPAs can be used to further amplify the readout pulses for the next stage. The state vector of the spins is also stored in precisely the same way, but with the use of only a single fiber core, since only \textit{N} pulses are needed for encoding. However, on the output of the spin memory, a spatial $1 \times  \textit{N}$ splitter is used to fan out the state vector into \textit{N} spatial copies, to be paired with each column of the Ising matrix, to support parallel MAC operations; again, a DOPA can be used immediately after the output coupler to compensate the tap and fan-out losses.

For the compute, these outputs are sent to an optical chip consisting of \textit{N} parallel circuits, each of which performs a sequence of MAC operations followed by a nonlinear function. First, to perform the MACs, the two optical pulse trains—one coming from an Ising-matrix column and the other a fanned out copy of the state vector—are incident on a nonlinear-optical waveguide that performs sum-frequency generation (SFG)~\cite{reifenstein2023coherent} (see Figure~\ref{fig:Comparison}(e)). This process takes the two inputs, pulse-by-pulse, and produces a third optical pulse train at their sum frequency, with pulse amplitude given (up to a constant factor) by the product of the two input amplitudes~\cite{reifenstein2023coherent}. As a specific example, we can choose the Ising matrix to be encoded as 1300 nm pulses and the state vector to be encoded as 1560 nm pulses; SFG thus produces a sum-frequency pulse train at ~710 nm~\cite{reifenstein2023coherent}. Crucially, however, the output of this SFG process is then looped back onto itself in a recirculating loop, creating a one-pulse register that continually multiplies and accumulates pairs of input pulses~\cite{reifenstein2023coherent}. As shown in Fig.~\ref{fig:Comparison}(c), after \textit{N} MAC operations, a final, strong pulse (one of the auxiliary control pulses, say at time bin \textit{N}+1) at the 1300 nm input performs difference-frequency generation (DFG) against the accumulated pulse in the register, which ejects the value of the register via a pulse at 1560 nm, the same wavelength as the spin variables~\cite{reifenstein2023coherent}. Finally, after the pulse exits the register, a suitable nonlinear-optical process, such as saturated DOPA, produces a nonlinear distortion of the field, which can be chosen to match such operations as tanh, ReLU, or sigmoidal functions~\cite{li2023all}. The last step in the architecture is to perform a SerDes operation, which takes the parallel outputs across the \textit{N} spatial channels and serializes them again into a single spatial channel with \textit{N} pulses in time. Passive delay lines on each output can delay the output by multiples of the pulse-to-pulse duration, and then a passive $\textit{N} \times 1$ splitter can be used to recombine the results. The serialized data are again reinjected back to the memory cavity that stores the spin variables, completing one loop of the dynamical algorithm.

To obtain order-of-magnitude physical scales for the architecture, the most crucial part to analyze is the optical memory storing the matrix elements, since it is the most resource-intensive part. Furthermore, if low-loss multicore fiber can be engineered to reach 0.2 dB/km attenuation, the roundtrip loss of the fiber loop itself is only 5\%. If we assume an additional 5\% for outcoupling, then the dominant optical energy cost of the memory lies in replenishing 10\% roundtrip loss. To reach 50 dB of SNR (corresponding to 8-bit precision), the coherent amplitude of the circulating pulses should nominally correspond to 25 thousand photons, or about 4 fJ per pulse and 800 W total circulating power (over all \textit{N} fiber cores) at 1300 nm. We would also like to sustain this optical memory over a significant number of roundtrips – accounting for an additional 20 dB overhead for SNR degradation (sufficient for 1000 roundtrips in principle), the total power dissipated is about 8 kW, amounting to a total optical energy consumption for this memory unit of 40 fJ/MAC (for $10^{12}$ MACs). It is also worth noting that at 4-bit precision (26 dB nominal SNR), these numbers come down by a factor of 250. Because the memory unit for the state vector only requires a single fiber core, its analogous energy consumption is effectively negligible. There is an additional need to precompensate for a 1 x \textit{N} fanout of the state vector, but the available gain in a TFLN DOPA can be as high as 55 dB$/cm/W^{1/2}$ ~\cite{jankowski2021dispersion, jankowski2022quasi}, therefore, a 1 cm waveguide can make up the fanout loss with about 1 W of optical peak power, with1 ps pulses, corresponding to only about 200 mW average pump power. 

To analyze the MAC unit, we consider that the available length of TFLN waveguide in the accumulator cavity is about 400 $\mu$m and that the peak power of the weight pulses is nominally around 0.2 mW (assuming 5\% outcoupling on 4 fJ pulses and no additional losses). This means the conversion factor for the signal pulses (from the output of the state vector memory to the sum-frequency in the cavity) is about 0.001~\cite{reifenstein2023coherent, jankowski2022quasi, jankowski2021dispersion, jankowski2024ultrafast}, which is in fact well matched to the factors typically used to discretize dynamical algorithms; if a larger conversion is needed, additional DOPA gain with TFLN OPAs can be used at a modest cost. The nonlinear activations downstream of the MAC occur only once per loop per waveguide, meaning that their duty cycle compared to all the elements above is 1/\textit{N} and therefore, they can be neglected. The last element is the \textit{N}  $\times$ 1 splitter and associated preamplification for the fan-in and SerDes operation, but, as argued above for the 1 $\times$ \textit{N}  fanout, this is also a negligible cost.

\subsubsection*{Practical challenges in all-optical coherent spatiotemporal architecture:}
\textcolor{black}{To reach a million-node Ising machine, the architecture needs to operate at a challenging but not implausible optical clock rate of 200 GHz, to store one million time-multiplexed pulses in $\sim$1 km of fibre, which can further be plausibly phase-stabilised in an experiment \cite{li2025all,honjo2021100} The all-optical coherent spatiotemporal architecture employs nonlinear processes such as DOPO and SFG; therefore, it must maintain uniform nonlinear conversion to avoid amplitude inhomogeneity. In addition, due to the short optical pulses (fs to sub-ps) and the high repetition-rate signal ($\sim$5 ps), the architecture requires careful consideration of dispersion compensation and fibre loop synchronisation, respectively. In general, the system requires N parallel TFLN cores (individual SFG units) to calculate each component of the output vector. Realistic designs for these devices are approximately 0.25 $mm^2$, so a large-scale wafer (8 to 18 inches) would fit the right order of magnitude in the required number of devices, but the engineering challenges of achieving large-scale wafer uniformity and yield remain formidable. However, CMOS electronics, where dense integration is important to mitigate the energy-cost and heating issues associated with electrical interconnects between chips, an all-optical architecture, the cost of delocalizing compute across dozens to hundreds of “optical chiplets” could be nearly as low as that of a fully integrated system if high optical coupling efficiency between chips can be engineered.}

\section{Conclusion} 
\label{sec:conclusion}
\textcolor{black}{Beyond conventional binary, pairwise Ising machines, two complementary generalizations— higher-dimensional spin encodings and higher-order Ising machines—could expand the capabilities of photonic optimization hardware. Binary spin-encoded Ising machines can become trapped in local minima and therefore often require auxiliary mechanisms, such as feedback-induced spin flips or stochastic perturbations. By contrast, higher-dimensional systems can provide additional pathways through the energy landscape, potentially increasing the probability of reaching lower-energy configurations, and ideally the ground state. Higher-dimensional spin encodings, or hyperspins, such as two-dimensional XY spins and three-dimensional Heisenberg spins, enlarge the state space of each spin and can provide smoother pathways through the energy landscape \cite{calvanese2022multidimensional}. In photonics, higher-dimensional spin systems can be implemented by hierarchically encoding multiple spin states using multi-pulse sequences, in which several spins are jointly encoded to form a D-dimensional continuous spin multiplet. Alternatively, nonlinear polarization oscillations in third-order nonlinear media can encode three-dimensional spherical spin on the Poincaré sphere \cite{chiavazzo2025ising}.} 

\textcolor{black}{Higher-order Ising machines (three-body or beyond) could solve some optimization problems more resource-efficiently than conventional pairwise Ising machines\cite{bybee2023efficient, ahsan2026higher, de2025incorporate}, because the number of possible interactions increases with the interaction order. Several barriers remain, including computational precision, readout and discretization, hardware complexity, and encoding imperfections \cite{ahsan2026higher}. In addition, different orders of interaction can scale unevenly with continuous spin amplitudes, leading to imbalances and performance degradation. However, rescaling the spin amplitudes by the mean absolute value across all spins mitigates amplitude imbalance, improves performance, and is well-suited for hardware implementation \cite{de2025incorporate}.}

Photonic Ising machines are a promising computing architecture for combinatorial optimization and offer potential advantages in speed, parallelism, and energy efficiency. However, current photonic implementations remain limited in scalability, reconfigurability, time-to-solution, and solution quality, and are yet to address many practical applications. In this perspective, we reviewed three fundamental classes of photonic Ising architectures and discussed key challenges, associated with scalability, time-to-solution, reconfigurability, and
hardware optimization. We also outlined a roadmap towards large-scale on-chip, free-space and spatiotemporal photonic Ising processors, with the goal of moving beyond million-node systems. Several open questions must be addressed before photonic Ising machines can become practical optimization accelerators. First, how can the algorithms implicitly implemented by photonic Ising machines be improved while preserving the advantages of photonic hardware? Second, how can advances in material and device fabrication, including reconfigurable metamaterials, large-scale TFLN SLMs, and ultrafast optoelectronic components, enable ultra-low power consumption with effective thermal and noise management? Third, how can architectures be simplified using concepts such as hybrid multiplexing \cite{calvanese2022multidimensional} and brain-inspired hierarchical networks \cite{tait2014} to realize adaptive, large-scale photonic systems? Finally, how can analog, optoelectronic, and optical or electronic ASICs be co-integrated and co-packaged at large scale? Addressing these challenges could enable photonic Ising machines to become a useful tool for applications in drug discovery, materials identification, synthetic biology, and chemical discovery.

\section*{Acknowledgments}
N.A.K., A.A., N.R., and B.J.S. are supported by the Natural Sciences and Engineering Research Council of Canada (NSERC). B.J.S. is also supported by the Canada Research Chairs Program and a Sloan Research Fellowship. P.L.M. was supported by the National Science Foundation (award CCF-1918549) and a David and Lucile Packard Foundation Fellowship. A.A. thanks IIT Delhi for the NFSG fund project number MI02794.

\subsection*{Author Contributions} 
All authors contributed substantially to the writing and discussion of the content. All authors contributed to reviewing and/or editing the manuscript before submission.

\subsection*{Ethics declarations}
Competing interests: The authors declare no competing interests.

\subsection*{Data Availability}
Data are available on reasonable request.
\bibliography{sample}

\end{document}